\documentclass[manuscript]{acmart}

\AtBeginDocument{%
  }

\setcopyright{acmlicensed}
\copyrightyear{2026}
\acmYear{2026}
\acmDOI{XXXXXXX.XXXXXXX}

\acmConference[CSCW '26 Growing Up (and Old) with AI Workshop]{2026 Conference on Computer-Supported Cooperative Work and Social Computing Workshop}{October 10--14, 2026}{Salt Lake City, UT, USA}
\acmBooktitle{CSCW '26 Growing Up (and Old) with AI: Co-Constructing the Future for Family-Centered AI Workshop, October 10--14, 2026, Salt Lake City, UT, USA}
\acmISBN{XXX-X-XXXX-XXXX-X/XX/XX}
\acmPrice{15.00}

\usepackage{longtable}
\usepackage{geometry}
\usepackage{pdflscape}
\usepackage[table]{xcolor}
\usepackage{colortbl}
\usepackage{tabularx}
\usepackage{array}
\usepackage{ragged2e} 
\usepackage{float}
\usepackage{multirow,booktabs,makecell}
\usepackage{graphicx}
\usepackage{subcaption}

\begin{document}

\title[When AI ``Works,'' When Does Help Begin?]{When AI ``Works,'' When Does Help Begin?: Intergenerational Support Around Older Adults' LLM Usage}

\author{Hyehyun Chu}
\affiliation{%
  \institution{KAIST}
  \city{Daejeon}
  \country{Republic of Korea}}
\email{hyenchu@kaist.ac.kr}

\author{Yuri Lee}
\affiliation{%
  \institution{POSTECH}
  \city{Pohang}
  \country{Republic of Korea}}
\email{yurilee@postech.ac.kr}
\author{Yeon Su Park}
\affiliation{%
  \institution{KAIST}
  \city{Daejeon}
  \country{Republic of Korea}}
  \email{yeonsupark@kaist.ac.kr}

\author{Saelyne Yang}
\affiliation{%
  \institution{Carnegie Mellon University}
  \city{Pittsburgh, PA}
  \country{USA}}
  \email{saelyney@andrew.cmu.edu}

\author{Juho Kim}
\affiliation{
  \institution{KAIST}
  \city{Daejeon}
  \country{Republic of Korea}}
  \email{juhokim@kaist.ac.kr}

\affiliation{
  \institution{SkillBench}
  \city{Santa Barbara, CA}
  \country{USA}}
  \email{juho@skillbench.com}

\renewcommand{\shortauthors}{Chu et al.}





\begin{abstract}

LLMs are becoming part of everyday life, including for older adults (OAs). OAs often learn digital technologies with younger family members, who have traditionally served as ``warm experts'' providing trusted and personalized operational help. LLMs expand this role: family supporters may also help OAs judge appropriate uses, consider what information to disclose, assess the credibility of outputs, and decide when AI-generated advice is safe to act on. We conducted a formative qualitative study with six OAs and seven younger adults (YAs), using semi-structured interviews and scenario-based think-aloud activities. 
OA participants described using LLMs to lighten their recurring reliance on family, while preserving family as a selectively invoked support channel. However, because LLMs rarely produced visible operational breakdowns, YAs had limited signals for when support was actually needed. Instead, YAs relied on OAs' partial disclosures and negotiated intervention through general warnings and self-imposed action boundaries. As a result, family support often solved an immediate problem without leaving reusable calibration knowledge for future use. Based on these findings, we propose design implications for intergenerational LLM support (e.g., consentful help requests, learning-oriented family support that preserves OA task ownership).

\end{abstract}

\maketitle

\section{Introduction} \label{intro}


Older Adults (OAs) (aged 60 or above \cite{guo2017older, world2023improving}) rarely encounter new technologies, such as smartphones, alone. Family members, particularly children and grandchildren, are a common source of informal support for learning smartphones, internet services, and other digital tools \cite{tang2022grandma,tsai2017playing}. Such supporters are often described as \emph{warm experts}: trusted members of an OA's close social network who offer patient, personalized help within an ongoing relationship, rather than formal and task-bounded instruction \cite{olsson2018warm}. This relational support can create meaningful learning opportunities, enabling OAs to develop digital-media skills through guidance that is responsive to their everyday needs and circumstances \cite{hanninen2021warm}.


Large language models (LLMs), such as ChatGPT, possess the possibility to reshape intergenerational support. Recent work describes younger family members moving from \emph{warm experts} who teach operation toward \emph{warm stewards} who sustain emotional support while monitoring, interpreting, and managing AI-related risks \cite{warmsteward2026}. In practice, this may involve helping older adults (OAs) judge appropriate uses of AI, consider what personal information to share, verify uncertain outputs, and decide when AI-generated advice is safe to act on. Yet these judgments can be difficult even for younger helpers themselves. The everyday interactional consequences of this expanded support role remain underexplored.

To examine how this emerging stewardship is enacted in everyday interactions, we conducted a formative study with six OAs and seven younger adults (YAs), using semi-structured interviews and scenario-based think-aloud activities. We examine this tension through three questions: 
\begin{enumerate}
    \item \textbf{RQ1:} How does intergenerational support around older adults' LLM use happen?
    \item \textbf{RQ2:} What challenges exist during intergenerational support around older adults' LLM use? 
\end{enumerate}

Our findings both extend the warm-steward account and reveal a feedback problem. For OAs, risk-related warnings arrived without a stated reason, so OAs often read them as criticism rather than actionable guidance. While YAs, lacking visibility into OAs' subsequent sessions, had no way to confirm whether their interventions had actually changed OAs' practices. We contribute an account of intergenerational LLM support as relational boundary work, and identify design opportunities for intergenerational support that close this feedback gap while preserving OA control and autonomy \cite{liu2022autonomy}.

\section{Methodology}

We recruited six OAs, consisting of two males and four females, with ages ranging from 60 to 70 years (\textit{M}~=~64.5, \textit{SD}~=~3.83). We also recruited seven YAs, consisting of five males and two females, with ages ranging from 24 to 34 years (\textit{M}~=~27.4, \textit{SD}~=~3.26). Participant demographics are summarized in Appendix~\ref{appendix}.

We recruited the OAs through online community forums in South Korea centered around keywords such as \textit{senior} and \textit{retiree}, where OAs actively participate, promotional posters placed at welfare centers, and snowballing through YA participants. The inclusion criteria required participants to (1) be at least 60 years old~\cite{guo2017older, world2023improving}, (2) report using LLM chatbots (e.g., ChatGPT, Gemini), and (3) have experience giving or receiving family-based support related to LLM chatbot use. Each OA received KRW 50,000 ($\approx$ USD 38), and each YA received KRW 20,000 ($\approx$ USD 14) for participation. The higher OA compensation reflected the greater difficulty of recruiting eligible OA participants.

Two OA--YA pairs participated as matched dyads. We assigned each pair a shared numeric suffix (OA1--YA1 and OA2--YA2). We assigned the remaining participants independent identifiers (OA3--OA6 and YA3--YA7).

\subsection{Procedure and Analysis}

\subsubsection{Study Procedure}
Each session lasted approximately 60 minutes and consisted of three parts: (1) an opening interview, (2) a live scenario-based task, and (3) a family-help recall interview.

\paragraph{Part 1: Opening Interview}
The session began with questions about participants' AI practices, confidence in using LLMs, and strategies for verifying AI-generated information.

\paragraph{Part 2: Live Scenario-Based Task}
Participants then selected one recent or realistic health-related concern and one legal or regulatory concern and used an LLM chatbot on their own phone or computer while thinking aloud and sharing their screen with the interviewer. During this task, we observed participants' prompt construction and refinement, verification behaviors, use of prior knowledge, and help-seeking behaviors.

\paragraph{Part 3: Family-Help Recall Interview}
The session concluded with a reconstruction of a recent family-help episode. We asked OAs about requesting, accepting, or resisting family help, and YAs about helping, gatekeeping, and their sense of responsibility toward older relatives. Both groups reflected on their preferences regarding timing, context sharing, and desired support.

\subsubsection{Data Analysis}
We audio-recorded and transcribed all interviews. We conducted thematic analysis on the transcripts following Braun and Clarke~\cite{Braun01012006} to identify recurring patterns across participants' accounts. Two authors carried out an iterative coding process: they first independently coded transcripts from three participants in each group (OA and YA) to develop preliminary codes and an initial codebook, then reconciled discrepancies and refined code definitions through discussion. The resulting codebook was then applied independently by the two authors to the remaining transcripts. The research team resolved newly emerging codes and interpretive disagreements through team meetings.

\section{Results}
This section presents five findings organized around our three research questions, covering how OAs incorporate LLMs into later life, how intergenerational support around this use unfolds, and what challenges arise within that support.

\subsection{RQ1: How Intergenerational Support around OAs' LLM Use Happens}

\subsubsection{F1: OAs used LLMs to regulate dependence within family relationships.}
OAs used LLMs not just to acquire a new skill, but to strengthen their autonomy within family relationships. Many turned to LLMs instead of asking their children for help with IT or everyday questions, wanting to remain independent and to spare their children's time and attention. This did not mean OAs were rejecting family help. Rather, it let them decide for themselves which questions were worth bringing to family in the first place. For instance, OA6 mentioned, \textit{``My son is really busy, so if I can search and learn it myself, there's no need to keep asking him. I just try it on my own, and only ask when I really can't figure it out.''} LLMs also enabled some OAs to showcase newly acquired competence to their families rather than only receiving help, while family remained a channel where they felt safe honestly admitting what they didn't know. OAs described their own use of LLMs as modest, everyday, and low-stakes, even when the domains involved (health, housing, finance, tax) were objectively high-stakes. They saw this as different from how YAs used LLMs, which they imagined as more consequential. Although OAs wanted to broaden how they used technology, they described having limited opportunities and contexts, such as retirement, to learn and expand their use.


\subsubsection{F2: OAs' preferred form of help was not always the one provided.}
We identified four forms of family support (see Table~\ref{tab}): \textit{doing it for the OA}, \textit{brief verbal rules}, \textit{demonstration}, and \textit{use-case observation}. Across episodes, these forms differed in who performed the task, what was conveyed through the support, and what prompted the support episode. In \textit{doing it for the OA}, YAs completed the task and handed back the result. \textit{Brief verbal rules} consisted of short warnings or general advice, often in response to concerns about trust or accuracy. In \textit{demonstration}, OAs performed the task while YAs guided them through the process and explained a method they could apply again. Finally, \textit{use-case observation} occurred when OAs encountered a family member's use of AI and subsequently tried a similar use on their own. Among these forms, OAs most frequently wanted demonstration, but the support they received did not consistently take this form.

\begin{table}[h]
\centering
\caption{Four forms of family support and their task characteristics}
\label{tab:SupportForm}
\begin{tabular}{p{2.7cm}p{4cm}p{4cm}p{5cm}}
\toprule
\textbf{Form} & \textbf{Task Characteristic} & \textbf{Trigger} & \textbf{Illustrative Quote} \\
\midrule
Doing it for the OA & YA performs the task and hands back only the result; prioritizes output quality over replicability & Operational friction & OA2: \textit{``[My child] is the type to just do it. Not teach me how. So next time I need to order something, I still can't do it myself and have to ask again.''} \newline YA2: \textit{``Writing the prompt directly seemed more effective than teaching.''} \\
\addlinespace
Brief verbal rules & Short, abstract warnings offered without operational detail & Evaluative concern,  \newline no visible friction & OA5: \textit{``When I ask `why shouldn't I?', there's no concrete answer... it just ends with `you're depending on it too much,' so I stop asking my son.''} \newline YA1: \textit{``I just said something like `don't trust it too much.'.''} \\
\addlinespace
Demonstration & OA performs the task while YA guides; conveys a transferable method & Deliberate teaching moment & YA5: \textit{``I suggest that my parent not just close the screen after reading the AI response, but copy it down on paper, think about whether it's correct, ask other people, or look up news articles.''} \\
\addlinespace
Use-case observation (modeling) & OA observes YA's output and independently attempts; conveys capability, not method & Incidental exposure & OA6: \textit{``I asked my son, `Did you go there and take that photo?' He said no, he made it with AI... I tried it myself.''} \\
\bottomrule
\end{tabular}
\end{table}

\subsection{RQ2: Challenges in Intergenerational Support around OAs' LLM Use}

\subsubsection{F3: Support rarely accumulated into reusable applicable knowledge.}
The mismatch between the support OAs preferred and the support they received mattered because different forms of help left different resources for future use. \textit{Doing it for the OA} could resolve an immediate problem, but it often left OAs with an outcome rather than a method they could reproduce. 
\textit{Brief verbal rules} posed a different problem: warnings such as ``don't trust it too much'' communicated caution but often gave OAs little guidance about what specifically to check or how to act differently. In some cases, such warnings also shut down further discussion rather than opening an opportunity for learning.
By contrast, demonstration provided OAs with strategies they could potentially reuse in later situations. However, YAs often selected support based on what seemed efficient and low-friction in the immediate interaction, rather than on what the OA could later reproduce independently. As a result, family support could successfully resolve individual episodes without accumulating into the reusable knowledge that OAs said they wanted for more independent future use.

\subsubsection{F4: Partial visibility turned support into unspoken boundary work.}
OA and YA each judged the other's ability and risk based on partial, fragmentary observations rather than a full picture of how the other actually used LLMs. YAs felt responsible for judging and protecting their parents' use without being able to see all of it, while OAs inferred YA's competence mainly from age and occupation. For instance, OA3 said, \textit{``I really have no idea... I don't know at all how my child uses it. I think they probably use it in their own way (...) but I think they're able to use it appropriately for their age.''} Standards for ``using AI well'' also diverged within dyads. OAs judged themselves by prompting skill, while YAs judged parents by their ability to distinguish trustworthy answers from plausible-sounding ones, showing that AI competence spans distinct dimensions (operation, utilization, calibration) that the two generations weighed differently. OAs and YAs also drew task boundaries differently. Given the same open-ended scenario, YAs framed questions at a practical management level while OAs raised higher-stakes matters simply because these were the real problems in their current lives, and YAs treated the meaningful boundary as the point where an unverified answer turned into action rather than whether a high-stakes question was asked at all. Families rarely discussed this boundary explicitly: OA chose what to show, YA filled gaps with assumptions, and intervention arrived only as brief, after-the-fact warnings, with no one's authority to set the boundary ever made explicit.

\section{Design Implications}

Building on these findings, we propose design implications for systems that support families in providing LLM support.
 
\textbf{Make uncertainty and task stakes visible for reflection and support.} Because an LLM produces a fluent answer to almost any input, difficulties with utilization and calibration rarely surface the way an operational failure would. OAs can struggle with a task while nothing on screen looks broken. Systems could avoid letting ``an answer was returned'' stand in for ``this was used well.'' Instead, systems could surface moments of uncertainty, such as an unusually broad or repeated question, a claim that contradicts what the OA already knows, or a jump straight from answer to action, as recognizable signals for reflection or family support.
 
\textbf{Turn family support into actionable, reusable guidance.} Brief warnings (``don't trust it too much'') and prompt completion each break down along a different axis: the former lacks the specificity to be acted on, the latter lacks the durability to be reused. Systems could help family support persist as concrete steps, attached to a reason, in a form the OA can call on again later, rather than dissolving into a one-off fix or a vague rule.

Furthermore, supporting OAs in structuring questions and expanding the range of everyday LLM uses could help them develop more reusable strategies for future LLM use. Structuring a question is itself a way of pre-setting evaluation criteria, so support given upfront also lowers the verification burden downstream. Without a learning channel after retirement, or a context for repeated practice, this front-loaded support matters more than age itself. Use-case observation, where an OA tries something new simply after seeing what a family member made, without any explanation, was in practice the pathway that most reliably expanded an OA's range of use. This suggests that systems could support this kind of ambient exposure rather than relying only on explicit instruction.
 
\textbf{Let the OA control the scope, timing, and audience of support.} OAs were not motivated by eliminating dependence on family, but by regulating what they handled alone and what they brought to family on their own terms. Systems could let the OA decide which conversations to share, whether to request an answer or just a verification method, and when to open the door to intervention. This sharing could work as more than a channel for requesting help. It could also let the OA show competence to family, reflecting the pride and common ground that some OAs described finding through AI use.



\bibliographystyle{plain}
\bibliography{reference}


\appendix
\section{Appendix} \label{appendix}

\begin{table}[htbp!]
  \caption{Demographic characteristics and self-reported LLM usage patterns of the six older-adult (OA) participants, including the specific LLM services used, length of usage experience, and typical use frequency.}
  \label{tab:OAparticipants}
  \centering
  \small
  \begin{tabular}{@{}ccclcc@{}}
    \toprule
    ID & Age & Gender & LLM services & Experience & Use frequency \\
    \midrule
    OA1 & 60 & F & Perplexity, Gemini & 3--6 months & 3--4 days/week \\
    OA2 & 61 & M & ChatGPT, Gemini & 6--12 months & 1--2 days/week \\
    OA3 & 63 & F & ChatGPT, Gemini & Over 1 year & Almost daily \\
    OA4 & 66 & M & ChatGPT, Gemini & 3--6 months & 1--2 days/week \\
    OA5 & 67 & F & Gemini & 3--6 months & Almost daily \\
    OA6 & 70 & F & Gemini & Over 1 year & 1--2 days/week \\
    \bottomrule
  \end{tabular}
\end{table}

\begin{table}[htbp!]
  \caption{Demographic characteristics and family context of the seven younger-adult (YA) participants, showing each participant's relationship to their older relative, whether that relative was also an OA participant in this study (indicated in the ``Matched OA'' column; ``---'' denotes older relatives who did not themselves participate), the older relative's age, and the pair's living arrangement and contact frequency.}
  \label{tab:YAparticipants}
  \centering
  \small
  \begin{tabular}{@{}ccclccl@{}}
    \toprule
    ID & Age & Gender & Relation & Matched OA & Relative age    &  Co-residence \& contact \\
    \midrule
    YA1 & 28 & M & Child & OA1 & 60 & Co-resident \\
    YA2 & 27 & M & Child & OA2 & 61 & Co-resident \\
    YA3 & 26 & M & Grandchild & --- & 79
      & Separate; contact 1--3 times/month \\
    YA4 & 24 & M & Grandchild & --- & 75
      & Separate; contact at least weekly \\
    YA5 & 25 & F & Child & --- & 64 & Co-resident \\
    YA6 & 34 & M & Child & --- & 62
      & Separate; contact at least weekly \\
    YA7 & 28 & F & Child & --- & 61 & Co-resident \\
    \bottomrule
  \end{tabular}
\end{table}

\end{document}